# Space-time Variant Self-growing Bandgap in Nonlinear Acoustic Metamaterial

Xin Fang[*], Jihong Wen[†], Dianlong Yu

*Laboratory of Science and Technology on Integrated Logistics Support, College of Intelligent Science, National University of Defense Technology, Changsha, Hunan 410073, China.*

Material band structure is key foundation for various modern technologies, but it was regarded as a space-time invariant feature. Acoustic metamaterials show extraordinary properties for processing elastic waves, but conventional realizations suffer from narrow bandgaps. Here we first report a nonlinear acoustic metamaterial whose band structure self-adapts to the propagation distance/time and the bandgap exhibits a self-growing behaviour stemming from giant nonlinear interaction. This space-time self-modulating characteristic highlights an unconventional understanding of the band structure, and the self-growth generates an ultralow and ultrabroad bandgap that breaks through the limitation of the mass law for linear locally resonant bandgaps. We also elucidate the self-adaptive mechanisms. This first demonstration sheds light on conceiving advanced devices and metamaterials with broadband, space-time variant bandgaps for wave self-manipulation.

Modern technologies rely on the band structures of materials to control electrons, photons and phonons [1, 2]. Advanced functions can be realized by tuning the bands of crystals [3-5]. However, band structures and bandgaps of linear periodic media are independent of space and time with respect to wave propagation, i.e., they are unique and invariable—the conventional opinion over the century since the band theory was proposed [6,7]. The uncontrolled adaptive modulations of wave propagation (i.e., self-modulation) realized by nonlinearity shows its great reliability in many applications [8]. Plenty of novel lasers and sensors have been created by the self-focusing [9, 10] or self-trapping [11] in nonlinear optical and acoustic media. Although the bandgaps of nonlinear media are amplitude-dependent, they are deemed to be invariable for specified parameters [12-14]. Actually, if the bandgaps could be self-adaptive, the devices relying on them would present interesting adaptive functions, but the feasibility remains unclear.

Moreover, low-frequency or broad bandgaps are desired in expansive applications [15,16]. Acoustic metamaterials [17-19] (AMs) offer unusual functions in manipulating low-frequency elastic waves [20-22]. However, in locally resonant (LR)-type linear AMs (LAMs) [23,24] widely studied since 2000, LR bandgaps are narrow in nature [24]. Obtaining broad LR bandgaps whose generalized width $\gamma>1$ remains a great challenge [25] due to the limitation of the mass law [26]. Nonlinearity helps make great advances [27-30], and nonlinear AMs (NAMs) [12,13,31] can boost the exploration of new features such as the harmonic generation [32], broadband chaotic mechanisms [33,34] and bridging coupling [35]. Bringing giant nonlinearity in NAM may generate new underlying physics for band structures and bandgaps, especially the self-modulating feature and broadband mechanism.

In this Letter, we report a NAM with giant nonlinear interactions which presents the first theoretical and experimental demonstration of the space-time variant self-growing bandgap and self-adaptive band structure. Its band structure self-adapts to the propagation distance/time, and the bandwidth increases spontaneously. Moreover, a self-grown ultralow and ultrabroad (double-ultra) bandgap that breaks through the limitation of the mass law evolves within a short distance. We elucidate its mechanisms with an equivalent approach.

*Model.*— The unusual phenomenon occurs in the NAM contains strongly coupled resonators featuring giant nonlinearity. A typical fundamental NAM prototype we fabricated is shown in FIG. 1. In the metacell, the primary oscillator $m_0$ is a rectangular bulk, and two neighbouring $m_0$ are coupled through a pair of springs whose entire stiffness is $k_0$. The hollow cylinder $m_2$ has a hole whose radius is 4.045 mm. $m_2$ couples to $m_0$ by two springs with total stiffness $k_2$. A steel sphere $m_1$ is set at the centre point of the cylindroid cavity of $m_2$, and the two are connected by a curved spring whose stiffness is $k_1$. There is a symmetrical clearance, $\delta_0=45\pm15$ μm, between the sphere and the cylinder wall at rest. When $|p(t)|>\delta_0$, $m_1$ will contact $m_2$, which leads to collisions—a strong nonlinear interaction. Therefore, $m_1$ and $m_2$ are vibro-impact oscillators, and the force between them, $F_N(t)$, is a piecewise function. The tiny clearance, $\delta_0$, is paramount to generating giant nonlinearity [36]. The NAM prototype consists of 30 metacells with the lattice constant $a=27$ mm. Its left terminal is connected to a vibration exciter through $k_0$, and the other end is fixed (FIG. 1c).

---

[*] xinfangdr@sina.com
[†] wenjihong@vip.sina.com

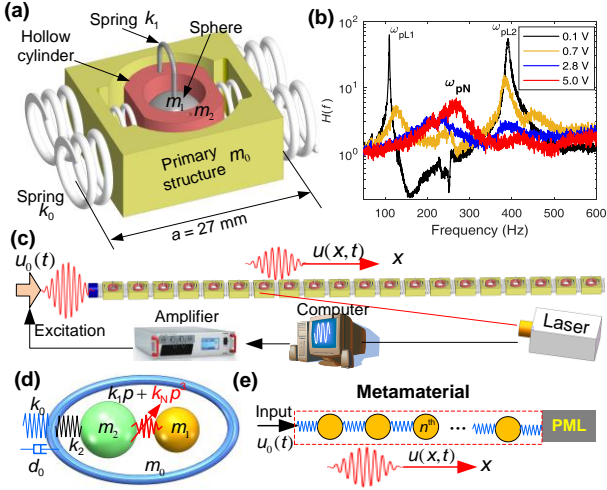

**FIG. 1 Prototype.** (a) Metacell structure. (b) Experimentally measured transmission of a single cell by driving $m_0$ with white-noise signals. Herein, the voltages, $v$ V, represent different driving levels; the result for 0.1 V denotes the linear state; $\omega_{pLi}$ and $\omega_{pN}$ symbolize peak frequencies for linear and nonlinear models, respectively, $i$=1, 2. (c) NAM chain consists of 30 meta-cells (only 20 cells are shown here). More experimental images are provided in Ref.[37]. (d) Theoretical triatomic metacell. (e) Numerical NAM model.

1d. The motion equations for the the $n^{th}$ cell are

$$\begin{cases} m_0 \ddot{u}_n = k_0(u_{n+1}+u_{n-1}-2u_n) \\ \qquad + d_0 k_0 (\dot{u}_{n+1}+\dot{u}_{n-1}-2\dot{u}_n) + k_2(z_n-u_n) \\ m_1 \ddot{y}_n = -F_N(t), \quad m_2 \ddot{z}_n = -k_2(z_n-u_n)+F_N(t) \end{cases} \quad (1)$$

Here, $u_n$, $y_n$, and $z_n$ denote the barycentre displacements of $m_0$, $m_1$, and $m_2$ in the $n^{th}$ cell, respectively, and $p=y-z$. We specify $F_N(t)=k_1 p+k_N p^3$ in theories for generality, where $k_1$ ($k_N$) denotes the linear (nonlinear) stiffness coefficient. Natural frequencies of individual oscillators are $\omega_i=2\pi f_i=\sqrt{k_i/m_i}$ $i$=0, 1, 2. If necessary, the damping coefficient $c_0=d_0 k_0$ in the primary oscillator is considered. As shown in FIG. 1e, the numerical NAM model consists of 120 periodic metacells. An optimized perfect match layer (PML) is connected to its terminal [37]. The incident wave $u_0(t)$ is launched from the other terminal.

Parameters in simulations and experiments are: $m_0$=5.8, $m_1$=2.1, $m_2$=2 g; $f_0$=322, $f_1$=100 and $f_2$=390.6 Hz. For the LAM with $k_N$=0, there is a Bragg bandgap and two LR bandgaps, LR1 and LR2 (FIG. 2a,c). Their ranges are LR1 (96.1, 110.4) Hz, with a generalized width $\gamma_1$=0.15, and LR2 (373.4, 462) Hz, with $\gamma_2$=0.24. Thus, LR1 and LR2 are narrow bandgaps. Moreover, if $k_1 \to \infty$, the triatomic AM degenerates to a diatomic AM characterized by a bandgap LRc (254.4, 356.4) Hz near $f_c = \sqrt{k_2/m_c}/2\pi$ =272.8 Hz, $m_c=m_1+m_2$, $\gamma$=0.4 (FIG. 2b,c).

If only the first linear resonances of $m_0$, $m_1$ and $m_2$ in the $x$ direction are considered, the entire metacell is equivalent to the triatomic configuration shown in FIG.

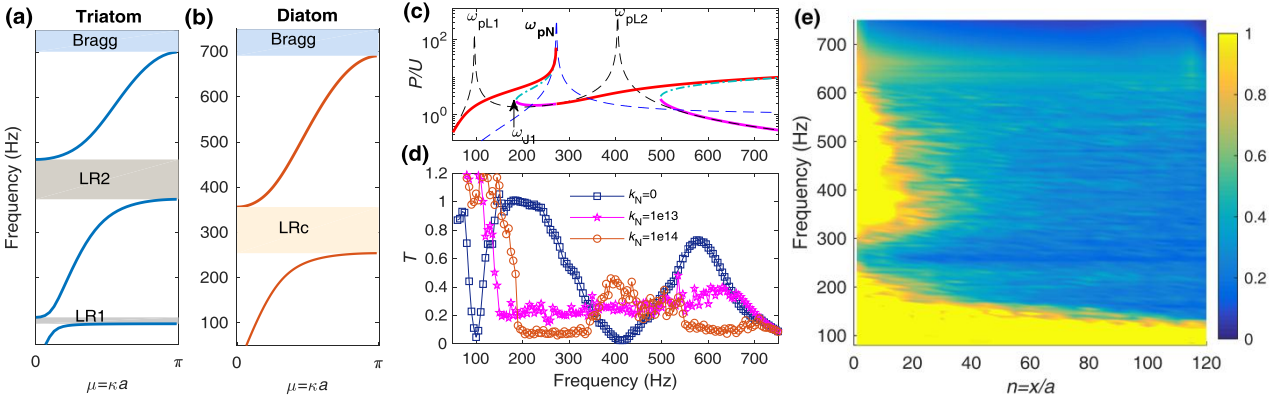

**FIG. 2** (a) Dispersive curves of the triatomic LAM. (b) Dispersive curves of the limiting diatomic LAM. The shading regions represent bandgaps. (c) Transmission of local resonances, i.e, the value $P/A$ solved with Eq. (4) below, where the black (blue dashed) line corresponds to the linear triatomic (diatomic) model; other curves are relevant to the cubic nonlinear triatomic model for $A$=5 μm. (d) Transmissions at the 100$^{th}$ cell, $T_{(100)}$. (e) Wave transmission $T(x=na, f_e)$. $A_0$=30 μm and $k_N$=1×10$^{13}$ N/m$^3$ are specified in (c-e).

*Phenomena.*—We take $k_N$=1×10$^{13}$ N/m$^3$ to show the nonlinear phenomena in theory. When nonlinearity appears, LR1 and LR2 become nonlinear LR bandgaps, and the coupling between them is visually defined as the bridging coupling [35]. The nonlinear strength for the cubic nonlinear system is $\sigma=3k_N A^2/k_1$. $\sigma$=32.6 for $A$=30 μm that indicates giant nonlinearity [36]. In experiments, the transmission spectrum, $H(f)$, of a single cell is measured by inputting different levels of white noises, as illustrated in FIG. 1b. By increasing the excitation, the first resonance notably shifts upwards at first, and the two resonances ultimately merge into a single resonance near 270 Hz. This tremendous shift verifies the giant nonlinearity generated by the clearance.

To show the wave transmission in the NAM, the sinusoidal wave packets, $u_0(t)=A_0\{[1-\cos(\omega t/10)]\sin\omega t\}$ for $t \leq 20\pi/\omega$, are adopted as input signals, where $\omega=2\pi f_e$ is the central frequency and $V_0 \approx 2\pi f_e A_0$. Wave transmission at the coordinate $x$ for $f_e$ is $T(x, f_e) = A_{max}(x, f_e)/A_0(f_e)$ where $A_{max}=\max[u(t)]$. At the $n^{th}$ cell, $T_{(n)}(f_e)=T(x=na, f_e)$.

For an infinite LAM, $T≈1$ in passbands in theory, and bands for $T≤0.2$ (but not $T→0$) are determined as bandgaps due to the broadband energy in the input sinusoidal packet. As shown in FIG. 2d, a surprising phenomenon is that the far-field transmission of the NAM is greatly reduced ($T_{(100)}≈0.2$) in the range 150-550 Hz. It is the bandgap at the 100$^{th}$ cell and its $\gamma$ reaches 2.67, a double-ultra bandgap. Over 550-670 Hz, $T_{(100)}≈0.3$ is much smaller than that of LAM. Regulars remain consistent for stronger nonlinearity. However, present studies fail to reveal the mechanism for this phenomenon.

The transmission in FIG. 2e illustrates that the near-field ($x<20a$) bandgap is not LR1 or LR2. Instead, it is LRc, indicating that $m_1$ and $m_2$ in the near-field meta-cells behave as an integrated resonator, $m_c$, due to the giant nonlinearity. Moreover, as $x$ increases, the initial frequency for the greatly attenuated band, $f_{st}$, shifts downwards gradually, and the transmission above LRc decreases much faster. Thus, the total band for wave suppression broadens until the entire range from LR1 to LR2 (even all above LR1) obtains bandgap effects. This process shows that the bandgap in this NAM adaptively grows from a narrow one, LRc, into a double-ultra one covering the range LR1-LR2, $\gamma=3.5$ (at least), which far exceeds the bandwidth of the LAM with $m_c/m_0=0.7$ [26]. We term this phenomenon a *self-growing bandgap*. As $x=c\ t$, where $c$ and $t$ denote the wave velocity and propagation time, a space-dependent property is also time-dependent.

Self-growth highlights a space-time self-modulating band structure, which modifies the conventional concept that the bandgap is invariable for specified condition. Furthermore, as the band structure determines the material properties, the wave velocity, refraction, focusing, scattering, transmission, insulation and attenuation that depend on the properties may be adaptive. Therefore, the new fundamental law enlightens us the extensive potential applications for manipulating elastic waves with broadband self-adaptive properties.

*Experiments.*—NAM responses under three driving levels, 2, 4 and 8 V, are measured by a Doppler laser vibrometer (FIG. 3). $A_0$ decreases with $f_e$ in the experiments (FIG. 3a). At a certain propagation distance, a greater transmission loss in 200-350 Hz is induced by a larger $A_0$ (FIG. 3b). A comparison between the numerical and experimental results of a diatomic LAM shows that there is an unavoidable weak damping, $d_0=4\times10^{-5}$ s, in the experiments. Moreover, to reproduce the experimental phenomena, we establish a numerical model consists of 30 cells containing clearance-nonlinearity, which confirms the credibility of experiments [37]. As shown in FIG. 3c, the near-field transmission exhibits a jump at a frequency, $f_B$. It is the grazing bifurcation [38] frequency of the vibro-impact oscillation [37].

Figures 3c,d indicate that the near-field ($n<3$) transmission near 273 Hz is greatly reduced, but waves in 0-255 Hz and 350-$f_B$ can still propagate. This result demonstrates that the near-field NAM behaves as a quasi-linear diatomic model that leads to the attenuation in LRc. As the propagation distance/time increases, waves in the near-field passbands, 200-$f_B$, are attenuated gradually due to the bandgap effect. The entire bandgap becomes broader at longer propagation distances, successfully demonstrating the space-time variant adaptive band structure and self-growth. The nonlinear bandgap at the 13$^{th}$ cell has reached 200-500 for levels 8 V (180-500 Hz for level 4 V [37]). Our experiments show that the bandgap grows into a double-ultra one within 9~13 cells that is much faster than the numerical prediction in FIG. 2e.

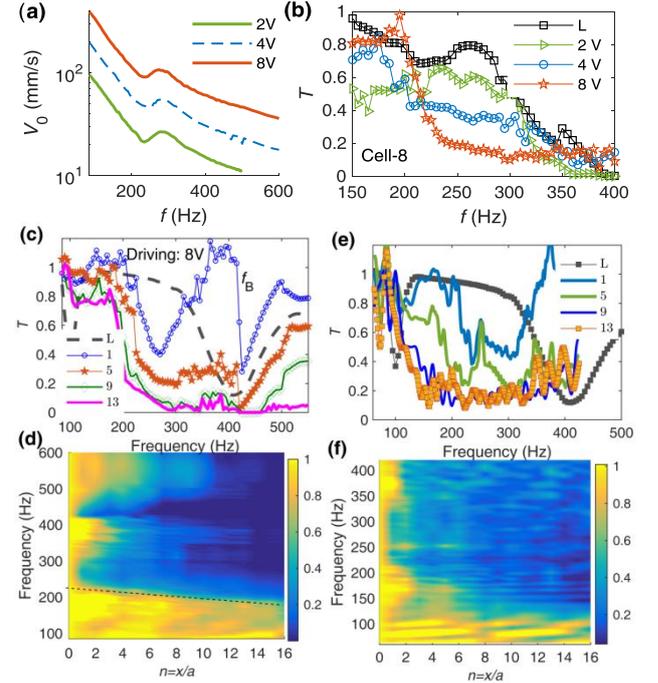

**FIG. 3 Experimental demonstration.** Panels (**a-d**) are results for inputting sinusoidal packets. (**a**) $A_0≈V_0/\omega$, where $V_0$ denotes the directly measured velocity amplitude. (**b**) Experimentally measured transmissions at 8$^{th}$ cell. Here, 2 V, 4 V and 8 V are the three driving levels for the NAM; 'L' represents the result for the diatomic LAM prototype. (**c**, **d**) Transmissions under 8 V. (**e**, **f**) Transmissions obtained in sweep-frequency experiments. In (**c**, **e**), the number, $n$, in legends represents the $n^{th}$ cell; 'L' represents the numerical result of the 5$^{th}$ cell in the linearized model. (**d h**) Transmission $T(x=na, f_e)$, $n=x/a$.

Moreover, this experiment still prove that opening the bandgap by observing the residual amplitude below 180 Hz requires a longer propagation distance, which is attributed to the challenge in attenuating all broadband energy of packets. The fast sweep-frequency experiment [37] (see FIG. 3e,f) shows a faster self-growing speed, especially for $f<250$ Hz. Herein the entire bandgap successfully expands to 130 Hz at the 13$^{th}$ cell, which is close to LR1. Certain peaks at high frequencies correspond to the standing waves arising from the boundary reflection of the finite NAM. Results demonstrate that the total bandgap can adaptively grow to at least 130~500 Hz ($\gamma$=2.85) within 9~13 cells. For $m_c/m_0$=0.7, this unusual double-ultra bandgap extends far beyond the mass law of a LR bandgap.

*Mechanisms.*—We propose an equivalent linearized approach based on the bifurcation of the coupled nonlinear local resonance to clarify the self-growing mechanisms [37]. A plane wave in the 1D medium is $u(x,t) = Ae^{i(\kappa_r x - \omega t)}$, $A(x) = A_0 e^{-\kappa_I x}$, where $A_0$ is the source amplitude; the wave number $\kappa = \kappa_r + i\kappa_I$, and $\mu = \kappa a = \mu_r + i\mu_I$. The dispersion equation of this NAM is

$$\cos\mu = 1 - \frac{m_0\omega^2}{2k_0} - \frac{\omega^2\omega_2^2[m_1\omega_{1eL}^2 + m_2(\omega_{1eL}^2 - \omega^2)]}{2k_0[m_2(\omega^2 - \omega_{1eL}^2)(\omega^2 - \omega_2^2) - m_1\omega^2\omega_{1eL}^2]} \quad (2)$$

where $\omega_{1eL}$ denotes the equivalent nature frequency of $m_1$,

$$\omega_{1eL} = \sqrt{\frac{\omega_{J1}^4 - \omega_{J1}^2\omega_2^2}{(1 + m_1/m_2)\omega_{J1}^2 - \omega_2^2}} \quad (3)$$

here, $\omega_{J1}$ is the first bifurcation frequency of the nonlinear local resonances as labelled in FIG. 2c, and we determine it by the algebraic equations:

$$\begin{cases} \omega^2 m_1 Y = k_1 P + 3k_N P^3/4 \\ (k_2 - \omega^2 m_2)(Y - P) - \omega^2 m_1 Y = k_2 A \end{cases} \quad (4)$$

where $Y$ and $P$ denote the amplitudes of $y(t)$ and $p(t)$, respectively. Therefore, the band structure is determined by the wave amplitude $A$. When considering damping, we replace $k_0$ in Eq. (2) by $k_0(1+i\omega d_0)$.

The bandgap can be described by $\kappa_I = \mu_I/a > 0$ solved with Eq. (2). The amplitude reduction in propagation distance $\Delta x$ denotes $e^{-\kappa_I \cdot \Delta x}$. Bandgaps lead to great attenuation attributing to the large $\kappa_I$. The amplitude-dependent regime in FIG. 4a shows that the band structure approaches the state of the diatomic LAM if giant nonlinearity is generated when $A$ is large, which agrees with the regulars in FIG. 2e. Furthermore, if $A$ decreases from a large value, LRc approaches LR1, and LR2 first decouples from the Bragg gap and then shifts to lower frequencies. A narrow band referred as *leakage band* is never swept by the adaptive bandgap. It results in the transmission peak near 370 Hz in FIG. 2d.

Moreover, we establish a large numerical model to separate the incident and reflected waves in the NAM [37]. Wave inside NAM is a superposition of the forwardly transmitting and backward reflected waves. If $(A_0, \omega)$ first appears in LRc, as represented by the spectra of 270 Hz in FIG. 4b, most input energy is reflected by several metacells near the incident boundary, thereby the amplitude decreases rapidly.

If $(A_0, \omega)$ first appears in passbands, $\kappa_I = 0$ for $d_0 = 0$ in linear regimes, thereby the amplitude never attenuates, and the band structure never varies in linear case. However, as represented by 200 Hz in FIG. 4c,d, waves transmitting into NAM undergo the harmonic generation [32] and chaotic responses that pump the fundamental energy to a broad band [37]. This process also reduces the total amplitude. Then after a certain distance, two effects appear. (i) The amplitude-dependent band structure varies and the bandgaps shift downwards; if a shifted bandgap covers $\omega$, a significant reduction, $e^{-\kappa_I \cdot \Delta x}$, is induced. (ii) All bandgaps are active to reflect the broadband energy. Once the attenuation begins, band structure varies and broader bands are swept by bandgaps, then more energy is reflected, thereby larger reductions are induced. This is a self-strengthening process. Finally, the observed far-field bandgap is significantly broadened, i.e., the self-adaptive band structure and the self-growing bandgap are realized.

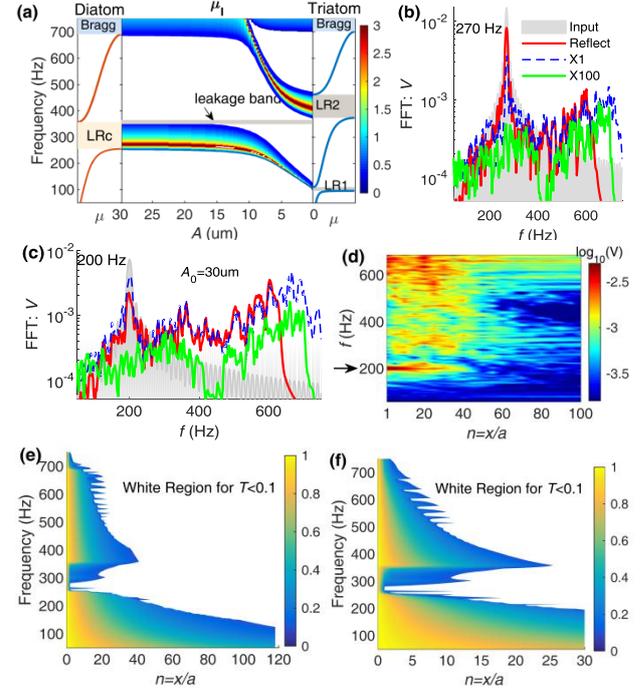

**FIG. 4** (**a**) Amplitude-dependent bandgap. The first and third insets are dispersion curves of the diatomic and triatomic LAMs. The second shading image is the distribution of the imaginary wave vector, $\mu_I = \kappa_I a$, in which the white regions represent $\mu_I = 0$ for passbands. (**b**, **c**) Velocity spectra for 270 and 200 Hz of the input wave, reflected wave, waves at 1$^{th}$ (X1) and 100$^{th}$ (X100) triatomic cells, respectively. (**d**) Distance-dependent velocity spectra for 200 Hz. (**e**, **f**) $T(x=na, f_e)$ solved with the equivalent method, $k_{dN}=2.5\times10^{-5}$. (**e**) $d_0=0$; (**f**) Damped case for $d_0=4\times10^{-5}$. In (c, d), $k_N=1\times10^{13}$ N/m$^3$, $A_0 = 30$ μm.

As analysed, without the damping, the self-growth is initiated by harmonic generation and chaotic responses. Defining $d_N$ as the entire attenuation rate induced by them, we have $A(x) = A_0 e^{-d_N x} e^{-\kappa_I x}$. Simulations and experiments indicate that $d_N \propto \omega$. We adopt $d_N = k_{dN}\omega$ to approximate this effect. By specifying $k_{dN}=2.5\times10^{-5}$, the analytical transmissions (FIG. 4e) agrees very well with the numerical illustration in FIG. 2e, which confirms the initializing regimes and verifies that the amplitude-distance-dependent bandgap is the main mechanism for the adaptive property. However, as $k_{dN}$ is small, opening the double-ultra bandgap needs 40 cells. Fortunately, a weak damping, $d_0=4\times10^{-5}$ s, can greatly accelerate the growth, as shown in FIG. 4e, and it agrees with the experiments. Thereby, damping is the reason for accelerating the self-growth.

*Conclusions.—* This letter presents the first theoretical and experimental demonstration of the time-space variant adaptive band structure and self-growing bandgap in a AM featuring giant nonlinear coupling. This finding changes the conventional opinion on band structures, and the bandwidth breaks through the limitation of the mass law. Due to the analogous properties of elastic, optical and electromagnetic waves, expanding the scope of this study is desirable.

[1] J. Kim, S. S. Baik, S. H. Ryu, Y. Sohn, S. Park, B. G. Park, J. Denlinger, Y. Yi, H. J. Choi, and K. S. Kim, Science **349**, 723 (2015).

[2] M. Abdi-Jalebi, Z. Andaji-Garmaroudi, S. Cacovich, C. Stavrakas, B. Philippe, J. M. Richter, M. Alsari, E. P. Booker, E. M. Hutter, and A. J. Pearson *et al.*, Nature **555**, 497 (2018).

[3] M. Schwarze, W. Tress, B. Beyer, F. Gao, R. Scholz, C. Poelking, K. Ortstein, A. A. Gunther, D. Kasemann, and D. Andrienko *et al.*, Science **352**, 1446 (2016).

[4] Z. Chen, X. Zhang, S. Lin, L. Chen, and Y. Pei, Natl Sci Rev (2018).

[5] R. Bogue, Sensor Rev **37**, 305 (2017).

[6] M. Born, and V. K. T. T, Z Phys, 297 (1912).

[7] R. Hofstadter, *Felix Bloch (1905-1983)* (National Academy of Sciences, Washington D.C., 1994).

[8] Z. Wu, Y. Zheng, and K. W. Wang, Phys. Rev. E **97**, 22209 (2018).

[9] X. Wang, H. Chen, H. Liu, L. Xu, C. Sheng, and S. Zhu, Phys. Rev. Lett. **119**, 33902 (2017).

[10] P. L. Kelley, Phys. Rev. Lett. **15**, 1005 (1965).

[11] S. Autti, Y. M. Bunkov, V. B. Eltsov, P. J. Heikkinen, J. J. Hosio, P. Hunger, M. Krusius, and G. E. Volovik, Phys. Rev. Lett. **108**, 145303 (2012).

[12] X. Fang, J. Wen, J. Yin, D. Yu, and Y. Xiao, Phys. Rev. E **94**, 52206 (2016).

[13] X. Fang, J. Wen, J. Yin, and D. Yu, Aip Adv **6**, 121706 (2016).

[14] R. K. Narisetti, M. J. Leamy, and M. Ruzzene, Journal of Vibration & Acoustics **132**, 31001 (2010).

[15] Y. Y. Chen, M. V. Barnhart, J. K. Chen, G. K. Hu, C. T. Sun, and G. L. Huang, Compos. Struct. **136**, 358 (2016).

[16] J. Y. Tsao, S. Chowdhury, M. A. Hollis, D. Jena, N. M. Johnson, K. A. Jones, R. J. Kaplar, S. Rajan, C. G. Van de Walle, and E. Bellotti *et al.*, Adv Electron Mater **4**, 1600501 (2018).

[17] Z. Liu, X. Zhang, Y. Mao, Y. Y. Zhu, Z. Yang, C. T. Chan, and P. Sheng, Science **289**, 1734 (2000).

[18] S. Zhang, C. Xia, and N. Fang, Phys. Rev. Lett. **106**, 24301 (2011).

[19] H. Abbaszadeh, A. Souslov, J. Paulose, H. Schomerus, and V. Vitelli, Phys. Rev. Lett. **119**, 195502 (2017).

[20] S. H. Lee, C. M. Park, Y. M. Seo, Z. G. Wang, and C. K. Kim, Phys. Rev. Lett. **104**, 54301 (2010).

[21] N. Kaina, F. Lemoult, M. Fink, and G. Lerosey, Nature **525**, 77 (2015).

[22] M. Yang, and P. Sheng, Annu Rev Mater Res **47**, 83 (2017).

[23] H. Ge, M. Yang, C. Ma, M. Lu, Y. Chen, N. Fang, and P. Sheng, Natl Sci Rev **5**, 159 (2018).

[24] G. Ma, and P. Sheng, Sci Adv **2**, e1501595 (2016).

[25] V. Romero-García, A. Krynkin, L. M. Garcia-Raffi, O. Umnova, and J. V. Sánchez-Pérez, J. Sound Vib. **332**, 184 (2013).

[26] The generalized width of a bandgap is $\gamma=(f_{cut}-f_{st})/f_{st}$, where $f_{st}$ ($f_{cut}$) is its initial (cutoff) frequency. A LAM's LR bandgap obeys the mass law $\gamma \approx \sqrt{1+m_c/m_0} -1$, where $m_c$ ($m_0$) denotes the mass of the local resonator and primary oscillator. A smaller mass ratio $m_c/m_0$ is better for most applications. For example, $\gamma \approx 0.3$ for $m_c/m_0=0.7$, and a huge value, $m_c/m_0=3$, is required to obtain $\gamma=1$ in LAM. Therefore, $\gamma>1$ can be regarded as a double-ultra bandgap because there is $\gamma<<1$ generally.

[27] G. Wehmeyer, T. Yabuki, C. Monachon, J. Wu, and C. Dames, Appl Phys Rev **4**, 41304 (2017).

[28] V. F. Nesterenko, C. Daraio, E. B. Herbold, and S. Jin, Phys. Rev. Lett. **95**, 158702 (2005).

[29] B. Liang, X. S. Guo, J. Tu, D. Zhang, and J. C. Cheng, Nat. Mater. **9**, 989 (2010).

[30] B. Deng, P. Wang, Q. He, V. Tournat, and K. Bertoldi, Nat Commun **9**, 3410 (2018).

[31] Y. Li, J. Lan, B. Li, X. Liu, and J. Zhang, J. Appl. Phys. **120**, 145105 (2016).

[32] X. Fang, J. Wen, D. Yu, G. Huang, and J. Yin, New J. Phys. **20**, 123028 (2018).

[33] X. Fang, J. Wen, B. Bonello, J. Yin, and D. Yu, Nat Commun **8**, 1288 (2017).

[34] X. Fang, J. Wen, B. Bonello, J. Yin, and D. Yu, New J. Phys. **19**, 53007 (2017).

[35] X. Fang, J. Wen, D. Yu, and J. Yin, Phys Rev Appl **10**, 54049 (2018).

[36] Nonlinear strength: The nonlinear strength σ<0.1 is weak nonlinearity and σ=0.3 can generate strongly nonlinear phenomena [12]. In our experiments, if we approximate the clearance nonlinearity with the smooth cubic nonlinearity, we obtain $k_N=3.3\times10^{12}$ N/m$^3$ [37]. The amplitude under 8 V reaches 533 and 53 μm for 100 and 300 Hz, respectively, and their σ=3392 and 33.5. They are huge numbers relative to σ=0.3. Therefore, it can generate giant nonlinear responses.

[37] see Supplementary Material.

[38] M. di Bernardo, C. J. Budd, A. R. Champneys, P. Kowalczyk, A. B. Nordmark, G. O. Tost, and P. T. Piiroinen, Siam Rev. **50**, 629 (2008).